\documentclass[12pt]{iopart}
\usepackage{amsthm}
\usepackage{graphicx}
\usepackage{setstack,cite}
\begin{document}
\date{}

\title[]
{Two-dimensional isochronous  nonstandard Hamiltonian systems}

\author{A. Durga Devi$^{1}$, R. Gladwin Pradeep$^{2}$, V. K. Chandrasekar$^{3}$,  and M. Lakshmanan$^{1}$}
\address{$^{1}$Centre for Nonlinear Dynamics, School of Physics,
Bharathidasan University, Tiruchirappalli - 620 024, India }
  \address{$^{2}$Department of Science \& Humanities, KCG College of Technology, Karapakkam, Chennai - 600 097, India }
\address{$^{3}$Centre for Nonlinear Science \& Engineering, 
School of Electrical \& Electronics Engineering, SASTRA University, Thanjavur 613 401, India }
%\date{\today}

\begin{abstract}
  We identify a generic class of two dimensional nonstandard Hamiltonian systems which exhibit isochronous behaviour.  This class of systems belongs to the two dimensional mixed Li\'enard- type equations and is obtained by generalizing the scalar modified Emden equation (MEE) to two dimensions.  We show that the generalized class of equations admits a Hamiltonian description and exhibits periodic and quasi-periodic oscillations for suitable choice of parameters and also $\emph PT$ symmetric property. 
%\keywords{Li\'enard oscillators \and periodic oscillations \and nonstandard Hamiltonian systems}
%\PACS{02.30.Hq \and 02.30.Ik \and 05.45-a}
\end{abstract}

\section{Introduction}
Certain nonstandard Lagrangian and Hamiltonian dynamical systems \cite{mukunda,arnold,Parra} encompass very interesting classes of nonlinear oscillators and admit fascinating dynamical properties \cite{musie,cari,jose} such as   isochronous oscillations,  linearization, nonlocal transformations and so on \cite{CSL:02,Nucci,cmee,ChML,Ch:02,EAAR,Chandru;jPA}.   In particular, the Li\'enard class of oscillators appear in the study of a wide range of fields such as seismology \cite{cartwright}, biological regulatory systems \cite{nicolas}, in the study of a self graviting stellar gas cloud \cite{shapiro}, optoelectronics, fluid mechanics \cite{kalashnik}.  Many of these Li\'enard class of equations admit limit cycle/periodic oscillations which are used to model many physical phenomenon.  Identifying such classes of coupled Li\'enard-type equations admitting isochronous oscillations is an interesting area of research.  Several procedures have been developed to construct and identify classes of isochronous oscillators.  In particular, Calogero \cite{Calogero:08} and Calogero and Leyvraz \cite{Calogero:08c,Calogero:08c1,Calogero:08c2,Calogero:08d,Partha:0a} have developed many techniques to generate isochronous oscillators.  In a recent paper a procedure to generate scalar isochronous systems recursively from a given Hamiltonian \cite{CDL:01} and a method to construct higher dimensional isochronous nonsingular Hamiltonian systems have  been discussed \cite{DGCL:01}.  In this paper we obtain a class of coupled Li\'enard type oscillator equations admitting isochronous oscillations by generalizing the nonstandard Lagrangian of the scalar system to coupled systems.  In order to do so let us consider the scalar Li\'enard-type linear/ quadratic and mixed type (in velocities) systems.  For example the Li\'enard equation with linear velocity term,
\begin{eqnarray}
\ddot x+F(x)\dot x+G(x)=0,\label{meee}
\end{eqnarray}
admits a class of interesting nonstandard type conserved Hamiltonian \cite{CSL:02} and isochronous solutions for the choice \cite{ChML},
\begin{eqnarray}
F(x)=3kx, \quad G(x)=k^2x^3+\omega^{2}x.\label{mee}
\end{eqnarray}
For the choice $\omega=0$ the above equation reduces to the modified Emden equation which is well studied in the literature and it occurs in the study of equilibrium configurations of a spherical cloud acting under the mutual attraction of its molecules and subject to the laws of thermodynamics \cite{Ch:02} and in the modelling of the fusion of pellets \cite{EAAR}. For the choice $\omega\ne0$ the above equation exhibits isochronous behaviour, that is it admits periodic oscillations with frequency of oscillation independent of the amplitude \cite{CSL:02}.
The general solution of this equation is given as 
\begin{eqnarray}
x(t)=\frac{A\sin(\omega t+\delta)}{1-\frac{kA}{\omega}\cos(\omega t+\delta)}.
\end{eqnarray}
The corresponding system admits a nonstandard Hamiltonian
\begin{eqnarray}
H=\frac{1}{2}\hat{F}(p)x^2+U(p), \label{Hammee}
\end{eqnarray}
with 
\begin{eqnarray}
\hat{F}(p)=\omega^2\bigg(1-\frac{2 k}{\omega^2}p  \bigg), \quad U(p)=\frac{\omega^{4}}{2k^2}\bigg(\sqrt{1-\frac{2k}{\omega^2}p}   -1\bigg)^2,\label{Hammee2}
\end{eqnarray}
where the canonically conjugate momentum
\begin{eqnarray}
p= \frac{\omega^2}{2k}\bigg(1-\frac{\omega^{4}}{(k\dot x+k^2x^2+\omega^2)^2}  \bigg) .
\end{eqnarray}

System (\ref{Hammee}) is \emph{PT} symmetric $(x\rightarrow-x$, $t\rightarrow-t$, $\dot {x}\rightarrow \dot {x})$ and is also exactly quantizable in momentum space \cite{chithi}.
The class of nonlinear oscillators containing both quadratic and linear terms in $\dot x$ is called a mixed Li\'enard-type equation.  It can be written in the form 
\begin{eqnarray}
\ddot x+J(x)\dot x^2+F(x)\dot x+G(x)=0,\label{lienard2}
\end{eqnarray}
where $F(x)$, $G(x)$ and $J(x)$ are functions of $x$.  For example with the choice
 \begin{eqnarray}
J(x)=\frac{f_x}{f},\quad F(x)=\frac{(r+2)h_x}{(r+1)f}, \quad G(x)=\frac{hh_x}{(r+1)f^2}, \,r\ne-1\label{func}
\end{eqnarray}
the system admits a nonstandard Lagrangian and Hamiltonian functions of the form
\begin{eqnarray}
L=\frac{1}{(f(x)\dot x+h(x))^r},\label{scalag}
\end{eqnarray}
where $f(x)$ and $h(x)$ are arbitrary functions of $x$, $r$ is a suitably chosen real positive parameter and the Hamiltonian
\begin{eqnarray}
H=\frac{p}{f}\bigg(\bigg(\frac{-r f}{p}\bigg)^{\frac{1}{r+1}}-h   \bigg)-\bigg(\frac{-r f}{p}  \bigg)^{\frac{-r}{r+1}},
\end{eqnarray}
with the conjugate momentum
\begin{eqnarray}
p=-rf(f\dot x+h)^{-(r+1)}.\label{scalagp}
\end{eqnarray}
In the above, $r$ is a real positive number so chosen such that $H$ is real and in (\ref{scalag})-(\ref{scalagp}) only the principal branch is taken when fractional powers appear.  Note that the MEE is a special case of the above system with  $f=1$, $r=1$ and 
 $h=kx^2$.  We also note here that this class of Lagrangian and Hamiltonian systems are also drawing considerable interest in connection with supersymmetry related partner systems \cite{curtright,bijan} and $\emph PT$-symmetric systems \cite{chithi}.  
 Now, it is of considerable importance to extend the study of above type nonstandard Lagrangian and Hamiltonian systems to higher degrees of freedom.  Particularly we wish to identify two dimensional nonstandard Hamiltonian systems which are isochronous and in future to study their exact quantization as in the case of MEE \cite{chithi}.  

A natural way of generalising the above one dimensional nonstandard Lagrangian is to modify 
(\ref {scalag}) to the form, 
 \begin{eqnarray}
L=\frac{1}{(f\dot x+g\dot y+h)^r}\label{arblag}
\end{eqnarray}
where $f=f(x,y), g=g(x,y), h=h(x,y)$.
In this case, however the Lagrangian turns out to be singular or degenerate \cite{Parra} 
which can be verified from the vanishing the Hessian,
\begin{eqnarray}
\Delta\equiv\left|
\begin{array}{ccc}
\frac{\partial^2 L}{\partial \dot{x}^2}&\frac{\partial^2 L}{\partial \dot{x}\partial \dot{y}}\\
\frac{\partial^2 L}{\partial \dot{y}\partial \dot{x}}&\frac{\partial^2 L}{\partial \dot{y}^2}
\end{array}
\right|
=0.
\end{eqnarray}
To obtain a nonsingular Lagrangian, a modification of the above Lagrangian with suitable terms is essential.  For this purpose, we make a simple minded extension of the above form (\ref{arblag}) judiciously and succeed to identify non-trivial coupled nonstandard and nonsingular Lagrangian type nonlinear evolution equations.  We then show how from the associated equations of motion one can obtain coupled mixed Li\'enard-type equations, which may be considered as the natural extension of the above single degree of freedom mixed Li\'enard-type equation (\ref{lienard2}).  Then the procedure is specifically illustrated for the MEE equation to obtain a two dimensional isochronous extension of the MEE equation.

The plan of the paper is as follows.  In section 2, we  introduce an  arbitrary form of nonsingular Lagrangian and use it to obtain the corresponding Newton's equation of motion.  However the resultant equation of motion is not in the  form of mixed Li\'enard-type class of oscillators.  By demanding the resultant dynamical equations are of generalized coupled mixed Li\'enard-type equations, we deduce the functional form of the allowed nonstandard Lagrangian.  In section 3, we show that the conditions obtained from the coupled mixed Li\'enard-type oscillators allow one to deduce another class of Li\'enard-type oscillators which exhibits periodic and quasiperiodic motions.  Also we have shown that the corresponding Hamiltonian obtained from the coupled nonsingular Lagrangian can be transformed into a two dimensional harmonic oscillator Hamiltonian through appropriate canonical transformations.  In section 4, we have shown that a special case of Li\'enard-type of oscillators leads to a two dimensional version of  the modified Emden equation which exhibits isochronous property.  
The general solution is shown to admit periodic as well as quasiperiodic solutions for suitable choices of parameters. Finally, in section 5, we present our conclusions.  

\section{Two dimensional coupled mixed Li\'enard-type equations}

We wish to identify a coupled mixed Li\'enard-type system which possesses a nonstandard Hamiltonian and admits isochronous solution by a suitable generalization of the scalar Lagrangian (\ref{scalag}) of MEE.  As noted earlier, the Lagrangian (\ref{arblag}) which is a natural generalization in two dimensions is singular.  One can overcome this problem by suitably redefining the Lagrangian in the form,

\begin{eqnarray}
L=\sum_{i=1}^{2}\frac{1}{(f_{i}\dot x+g_{i}\dot y+h_{i})^{r_{i}}}.\label{eq7}
\end{eqnarray}
where $f_{i}=f_{i}(x,y)$, $g_{i}=g_{i}(x,y)$, $h_{i}=h_{i}(x,y)$, $i=1,2$.  With this choice of the Lagrangian we can show that Hessian in the present case is non-zero,
\begin{eqnarray}
\Delta\equiv\left|
\begin{array}{ccc}
\frac{\partial^2 L}{\partial \dot{x}^2}&\frac{\partial^2 L}{\partial \dot{x}\partial \dot{y}}\\
\frac{\partial^2 L}{\partial \dot{y}\partial \dot{x}}&\frac{\partial^2 L}{\partial \dot{y}^2}
\end{array}
\right|
\ne0.
\end{eqnarray}
Consequently, the Lagrangian is nonsingular.
From the above Lagrangian, the equation of motion can be obtained as  
\begin{eqnarray}
&&\ddot x= \bigg(g_{2}(x,y)q_{1}(\dot x,\dot y,x,y)+g_{1}(x,y)q_{2}(\dot x,\dot y,x,y)  +  d_{1}(x,y)\dot x^2
\nonumber\\
&&\hspace{0.5cm}+d_{2}(x,y)\dot y^2+d_{3}(x,y)\dot x \dot y+ s_{1}(x,y)\dot x+s_{2}(x,y)\dot y+u_{1}(x,y)    \bigg),\label{seceq1}\\
%\end{eqnarray}
%\begin{eqnarray}
&&\ddot y= -\bigg(f_{2}(x,y)q_{1}(\dot x,\dot y,x,y)+f_{1}(x,y)q_{2}(\dot x,\dot y,x,y)+d_{4}(x,y)\dot x^2 
\nonumber\\
&&\hspace{0.5cm}+d_{5}(x,y)\dot y^2+d_{6}(x,y)\dot x \dot y+ s_{3}(x,y)\dot x+ s_{4}(x,y)\dot y+u_{2}(x,y)     \bigg),\label{seceq2}\\
%\end{eqnarray}
&&\hspace{-1cm}\mbox{where}\nonumber\\ 
%\begin{eqnarray}
&&q_{1}=\frac{(r_{2}+1)}{\Delta^{2}r_{1}}\bigg(r_{2}(f_{1}\dot x+g_{1}\dot y+h_{1})^{(r_{1}+2)} (f_{2}\dot x+g_{2}\dot y+h_{2})^{-(r_{2}+1)}\nonumber\\
&&\hspace{0.5cm}\times\bigg(f_{2}h_{2y}-g_{2}h_{2x}+(f_{2y}-g_{2x})(\dot x f_{2}+\dot y g_{2})\bigg)\bigg),\label{q1}\\
%\end{eqnarray}
%\begin{eqnarray}
&&q_{2}=\frac{(r_{1}+1)}{\Delta^{2}r_{2}}\bigg(r_{1}(f_{1}\dot x+g_{1}\dot y+h_{1})^{-(r_{1}+1)} (f_{2}\dot x+g_{2}\dot y+h_{2})^{(r_{2}+2)}\nonumber\\
&&\hspace{0.5cm}\times\bigg(f_{1}h_{1y}-g_{1}h_{1x}+(f_{1y}-g_{1x})(\dot x f_{1}+\dot y g_{1})\bigg)\bigg),\label{q2}
\end{eqnarray}
%\end{subequations}
%\begin{subequations}
%\begin{eqnarray}
%%%%%%right\begin{subequations}
\begin{eqnarray}
&&d_{1}=\frac{1}{\Delta^{2}}\bigg(f_{1}f_{2}(g_{1}(f_{2y}-g_{2x})(r_{1}+1)+g_{2}(f_{1y}-g_{1x})(r_{2}+1))\nonumber\\
&&\hspace{0.5cm}+(r_{1}+1)(r_{2}+1)(g_{1}g_{2}(f_{2}f_{1x}+f_{1}f_{2x})-f_{2}f_{2x}g_{1}^2-f_{1}f_{1x}g_{2}^2) \bigg),\\
%\end{eqnarray}
%\begin{eqnarray}
&&d_{2}=\frac{1}{\Delta^{2}}\bigg(g_{1}^2(r_{1}+1)(g_{2}(f_{2y}-g_{2x})-f_{2}g_{2y}(r_{2}+1))+g_{2}^2(r_{2}+1)\nonumber\\
&&\hspace{0.5cm}\times(g_{1}(f_{1y}-g_{1x})-f_{1}g_{1y}(r_{1}+1))+g_{1}g_{2}(r_1+1)\nonumber\\
&&\hspace{0.5cm}\times(r_{2}+1)(f_{1}g_{2y}+f_{2}g_{1y})   \bigg),\\
%\end{eqnarray}
%\begin{eqnarray}
&&d_{3}=\frac{1}{\Delta^{2}}\bigg(g_{1}g_{2}(f_{2}(r_{2}+1)( f_{1y}(r_{1}+2)+g_{1x}r_{1})+f_{1}(r_{1}+1)\nonumber\\
&&\hspace{0.5cm}\times(f_{2y}(r_{2}+2)+g_{2x}r_{2}))-g_{1}^2f_{2}(r_{1}+1)(f_{2y}r_{2}+ g_{2x}(r_{2}+2))\nonumber\\
&&\hspace{0.5cm}-g_{2}^2f_{1}(r_{2}+1)(f_{1y}r_{1}+ g_{1x}(r_{1}+2))  \bigg),\\
%\end{eqnarray}
%\begin{eqnarray}
&&d_{4}=\frac{1}{\Delta^{2}}\bigg(f_{1}^2(r_{1}+1)(f_{2}(f_{2y}-g_{2x})+ f_{2x}g_{2}(r_{2}+1))+f_{2}^{2}(r_{2}+1)\nonumber\\
&&\hspace{0.5cm}\times(f_{1}(f_{1y}-g_{1x})+ f_{1x}g_{1}(r_1+1))- f_{1}f_{2}(r_1+1)(r_2+1)\nonumber\\
&&\hspace{0.5cm}\times(f_{2x}g_{1}+f_{1x}g_{2}) \bigg),
\end{eqnarray}
\begin{eqnarray}
&&d_{5}=\frac{1}{\Delta^{2}}\bigg(g_{1}g_{2}(f_{1}(f_{2y}-g_{2x})(r_1+1)+f_{2}(f_{1y}-g_{1x})(r_2+1))\nonumber\\
&&\hspace{0.5cm}+(f_{2}g_{1}-f_{1}g_{2})(f_{2}g_{1y}-f_{1}g_{2y})(r_{1}+1)(r_{2}+1) \bigg), \\
%\end{eqnarray}
%\begin{eqnarray}
&&d_{6}=\frac{1}{\Delta^{2}}\bigg(f_{1}^2g_{2}(r_1+1)(g_{2x}r_{2}+f_{2y}(r_2+2))+f_{2}^2g_{1}(r_2+1)(g_{1x}r_{1}\nonumber\\
&&\hspace{0.5cm}+ f_{1y}(r_1+2))-f_{1}f_{2}(g_{1}(r_1+1) (f_{2y}r_{2}+g_{2x}(r_2+2))+g_{2}(r_2+1)\nonumber\\
&&\hspace{0.5cm}\times(f_{1y}r_{1}+ g_{1x}(r_1+2)))\bigg),\\
%\end{eqnarray}
%\begin{eqnarray}
&&s_{1}=\frac{1}{\Delta^{2}}\bigg(g_{1}(r_1+1)(f_{1}(h_{2}(f_{2y}-g_{2x})+f_{2}h_{2y})-f_{2}g_{1}h_{2x}(r_2+2))\nonumber\\
&&\hspace{0.5cm}+g_{2}(r_2+1)(f_{2}(h_{1}(f_{1y}-g_{1x})+f_{1}h_{1y})-f_{1}g_{2}h_{1x}(r_1+2))\nonumber\\
&&\hspace{0.5cm}+ g_{1}g_{2}(r_1+1)(r_2+1) (f_{2}h_{1x}+f_{1}h_{2x})   \bigg), \\
%\end{eqnarray}
%\begin{eqnarray}
&&s_{2}=\frac{1}{\Delta^{2} }\bigg(g_{1}^2(r_1+1)(h_{2}(f_{2y}-g_{2x})-g_{2}h_{2x}-f_{2}h_{2y}(r_2+1))\nonumber\\
&&\hspace{0.5cm}
+g_{2}^2(r_2+1) (h_{1}(f_{1y}-g_{1x})-g_{1}h_{1x}-f_{1}h_{1y}(r_1+1))+g_{1}g_{2}
\nonumber\\
&&\hspace{0.5cm}
\times(f_{2}h_{1y}(r_1+2)(r_2+1)+f_{1}h_{2y}(r_1+1)(r_2+2))  \bigg),\\
%\end{eqnarray}
%\begin{eqnarray}
&&s_{3}=\frac{1}{\Delta^{2}}\bigg(f_{1}^2(r_1+1)(h_{2}(f_{2y}-g_{2x})+f_{2}h_{2y}+g_{2}h_{2x}(r_2+1))\nonumber\\
&&\hspace{0.5cm}+f_{2}^2(r_2+1)(h_{1}(f_{1y}-g_{1x})+f_{1}h_{1y}+g_{1}h_{1x}(r_1+1))
\nonumber\\
&&\hspace{0.5cm}-f_{1}f_{2}(g_{2}h_{1x}(r_1+2)(r_2+1)
+g_{1}h_{2x}(r_1+1)(r_2+2))  \bigg),\\ 
%\end{eqnarray}
%\begin{eqnarray}
&&s_{4}=\frac{1}{\Delta^{2}}\bigg(f_{1}g_{1}(r_{1}+1)(h_{2}(f_{2y}-g_{2x})-g_{2}h_{2x})+f_{2}g_{2}(r_{2}+1)\nonumber\\
&&\hspace{0.5cm}\times(h_{1}(f_{1y}-g_{1x})-g_{1}h_{1x})-f_{1}f_{2}(r_{1}+1)(r_{2}+1)(g_{2}h_{1y}+g_{1}h_{2y})\nonumber\\
&&\hspace{0.5cm}+f_{1}^2g_{2}h_{2y}(r_{1}+1)(r_{2}+2)+f_{2}^2g_{1}h_{1y}(r_{1}+2)(r_{2}+1) \bigg),\\
%\end{eqnarray}
%\begin{eqnarray}
&&u_{1}=\frac{1}{\Delta^{2}}\bigg[g_{1}h_{2}(f_{1}h_{2y}-g_{1}h_{2x})(r_1+1)
+g_{2}h_{1}(f_{2}h_{1y}-g_{2}h_{1x})\nonumber\\
&&\hspace{0.5cm}\times(r_2+1)\bigg], \\
%\end{eqnarray}
%\begin{eqnarray}
&&
u_{2}=\frac{1}{\Delta^{2}}\bigg[f_{1}h_{2}(f_{1}h_{2y}-g_{1}h_{2x})(r_1+1)+f_{2}h_{1}(f_{2}h_{1y}-g_{2}h_{1x})\nonumber\\
&&\hspace{0.5cm}\times(r_2+1)\bigg], \\
%\end{eqnarray}
%\begin{eqnarray}
&&\Delta=(f_{2}g_{1}-f_{1}g_{2})\sqrt{(r_{1}+1)(r_{2}+1)}. 
\end{eqnarray}
%\endnumparts
%%%%%%%%%%right\end{subequations}

The mixed scalar Li\'enard-type equation  (\ref{lienard2}) has only linear and quadratic terms in $\dot x$.  On the other hand  the coupled system of second order equations of motion (\ref{seceq1}) and (\ref{seceq2}), obtained from the nonsingular Lagrangian (\ref{eq7}) is not in the class of mixed Li\'enard-type oscillators because it contains higher/different powers of $\dot x$ and  $\dot y$ than quadratic and linear powers.  By equating  these higher/different degree coefficients to zero, analyzing them and making use of the results in the original coupled equations (\ref{seceq1}) and (\ref{seceq2}), we can obtain the relevant evolution equations.  For this purpose we equate the terms $q_{1}$ and $q_{2}$ in equations (\ref{seceq1}) and (\ref{seceq2}) to zero as they contain higher-degree terms in $\dot x$, $\dot y$ and their products.  Therefore we take
\begin{eqnarray}
q_{1}(\dot x,\dot y, x ,y)=0, \quad q_{2}(\dot x,\dot y, x ,y)=0. \label{q1q2}
\end{eqnarray}
Solving equation (\ref{q1}) and (\ref{q2}),  we get a set of partial differential equations for the variables $f_{i}$ and $g_{i}$, $(i=1,2)$. 
We can easily see that from $q_{1}(\dot x,\dot y, x ,y)=0$, we get
\begin{eqnarray}
f_{2y}=g_{2x},\quad  f_{2}=\frac{g_{2}h_{2x}}{h_{2y}},\label{condition1}
\end{eqnarray}

Similarly from $q_{2}(\dot x,\dot y, x ,y)=0$, we get
\begin{eqnarray}
f_{1y}=g_{1x},\quad f_{1}=\frac{g_{1}h_{1x}}{h_{1y}} , \label{condition}
\end{eqnarray}

%and substituting $f_{1}$ and $f_{2}$ in $\Delta=(f_{2}g_{1}-f_{1}g_{2})$, then
%\begin{eqnarray}
%\Delta=\frac{1}{h_{1y}h_{2y}}(g_{1}g_{2}(h_{1y}h_{2x}-h_{1x}h_{2y})),\nonumber
%\end{eqnarray}
On substituting the above forms in equations (\ref{seceq1}) and (\ref{seceq2}), we obtain the coupled system of mixed  Li\'enard-type class of oscillators of the form
%%%%%right\begin{subequations}
 
\begin{eqnarray}
&&\ddot x=\frac{-1}{\hat{\Delta}g_{1}g_{2}r_{12}}\bigg[g_{1}g_{2}\bigg((r_{1}+1)(r_{2}+1)\bigg((f_{2x}g_{1}-f_{1x}g_{2})\dot x^2-(g_{1y}g_{2}-g_{1}g_{2y})\dot y^2\nonumber\\
&&\hspace{0.7cm}-2(g_{1x}g_{2}-g_{1}g_{2x})\dot x\dot y\bigg)   +(r_{1}+1)(r_{2}+2)(h_{2x}\dot x+h_{2y}\dot y)g_{1}\nonumber\\
&&\hspace{0.7cm}-(r_{1}+2)(r_{2}+1)(h_{1x}\dot x+h_{1y}\dot y)g_{2}\bigg)-g_{2}^2h_{1}h_{1y}(r_2+1)\nonumber\\
&&\hspace{0.7cm}+g_{1}^2h_{2}h_{2y}(r_1+1)    \bigg] ,  \label{eq91}     
\end{eqnarray}
\begin{eqnarray}
&&\ddot y=\frac{1}{\hat{\Delta}g_{1}g_{2}r_{12}h_{1y}h_{2y}}\bigg[g_{1}g_{2}
\bigg((r_1+1)(r_2+1)  \bigg((f_{2x}g_{1}h_{1x}h_{2y}-f_{1x}g_{2}h_{1y}h_{2x})\dot x^2\nonumber\\
&&\hspace{0.7cm}+(g_{2y}g_{1}h_{1x}h_{2y}-g_{1y}g_{2}h_{1y}h_{2x})\dot y^2+2(g_{1}g_{2x}h_{1x}h_{2y}-g_{1x}g_{2}h_{1y}h_{2x})\dot x\dot y\bigg)
\nonumber\\
&&\hspace{0.7cm}-(r_{1}+2)(r_{2}+1)g_{2}h_{1y}h_{2x}(h_{1x}\dot x+h_{1y}\dot y)\nonumber\\
&&\hspace{0.7cm}+(r_{1}+1)(r_{2}+2)g_{1}h_{2y}h_{1x}(h_{2x}\dot x+h_{2y}\dot y)\bigg)
+g_{1}^2h_{2}h_{1x}h_{2y}^2(r_1+1)\nonumber\\
&&\hspace{0.7cm}-g_{2}^2h_{1}h_{1y}^2h_{2x}(r_2+1)    \bigg]. \label{eq92}  
\end{eqnarray}
%%%right\end{subequations}
%where the various functions are defined in equation (\ref{q1q2}).
The Hamiltonian associated with (\ref{eq91}) and (\ref{eq92}) corresponding to the Lagrangian (\ref{eq7}) can now be written down as 

\begin{eqnarray}
&&H=\frac{r_{12}}{\hat{\Delta}}\bigg[\frac{g_{1}}{h_{1y}}(p_{2}h_{1x}-p_{1}h_{1y})\bigg(h_{2}-\bigg(\frac{g_{1}(h_{1x}p_{2}-h_{1y}p_{1})r_{12}}{h_{1y}r_{2}\hat{\Delta}}  \bigg)^\frac{-1}{r_2+1}\bigg)\nonumber\\
&&\hspace{0.7cm}  +\frac{g_{2}}{h_{2y}}(p_{1}h_{2y}-p_{2}h_{2x})\bigg(h_{1}-\bigg(\frac{g_{2}(h_{2y}p_{1}-h_{2x}p_{2})r_{12}}{h_{2y}r_{1}\hat{\Delta}}\bigg) ^\frac{-1}{r_{1}+1}\bigg)\bigg]\nonumber\\
&&\hspace{0.7cm}-\bigg(\frac{g_{1}(h_{1x}p_{2}-h_{1y}p_{1})r_{12}}{h_{1y}r_{2}\hat{\Delta}}\bigg)^\frac{r_{2}}{r_2+1}-\bigg(\frac{g_{2}(h_{2y}p_{1}-h_{2x}p_{2})r_{12}}{h_{2y}r_{1}\hat{\Delta}}\bigg)^\frac{r_{1}}{r_1+1},
\end{eqnarray}
where $\hat{\Delta}=g_{1}g_{2}(h_{1y}h_{2x}-h_{1x}h_{2y})(h_{1y}h_{2y})^{-1}r_{12},\,\,r_{12}=[(r_{1}+1)(r_{2}+1)]^{\frac{1}{2}}$.
Here the conjugate momenta $p_{1}$ and $p_{2}$ are defined as 
%%%%%%%right\begin{subequations}
\begin{eqnarray}
&&\hspace{-2cm}p_{1}=L_{\dot x}=-\frac{f_{1}r_{1}}{(f_{1}\dot x+g_{1}\dot y+h_{1})^{r_{1}+1}}-\frac{f_{2}r_{2}}{(f_{2}\dot x+g_{2}\dot y+h_{2})^{r_{2}+1}},\\
&&\hspace{-2cm}p_{2}=L_{\dot y}=-\frac{g_{1}r_{1}}{(f_{1}\dot x+g_{1}\dot y+h_{1})^{r_{1}+1}}-\frac{g_{2}r_{2}}{(f_{2}\dot x+g_{2}\dot y+h_{2})^{r_{2}+1}}.
\end{eqnarray}
%%%%%%%right\end{subequations}

\section{\bf{Reduction to a subclass exhibiting quasiperiodic motion}}
~~~~~~~~	In our further analysis, for simplicity, we assume the parameters $r_{1}$ = $r_{2}$ = $1$ in the Lagrangian given by (\ref{eq7}).  Now, let the quantities $(f_{i}\dot x+g_{i}\dot y)$, $i=1,2,$ be the total derivatives (when $r_{1}$ = $r_{2}$ = $1$) of certain functions $\rho_{i}(x,y)$.  
%On comparing  $(f_{i}\dot x+g_{i}\dot y)$ with the derivatives of the functions $G_{i}$,  we can identify the conditions, $f_{iy}=g_{ix}$, $i=1,2,$ 
\begin{eqnarray}
f_{i}\dot x+g_{i}\dot y=\frac{d}{dt}[\rho_{i}(x,y)]=\rho_{ix}\dot x+\rho_{iy}\dot y.\label{totder}
\end{eqnarray}
From the above equation, we find $f_{i}=\rho_{ix}$, $g_{i}=\rho_{iy}$, $i=1,2$.  Substituting Eq. (\ref{totder}) in Eq. (\ref{eq7}) we get
\begin{eqnarray}
L=\sum_{i=1}^{2}\frac{1}{(\rho_{ix}\dot x+\rho_{iy}\dot y+h_{i})}\label{modlag}.
\end{eqnarray}

%Comparing the right hand side of (\ref{totder}) with $(f_{i}\dot x+g_{i}\dot y)$, we have
%\begin{eqnarray}
%f_{iy}=g_{ix}=G_{ixy}
%\end{eqnarray}
%As $G_i(x,y)$ is simply an arbitrary function of $(x,y)$, it can be replaced by another arbitrary function, lets say $f_i(x,y)$.  Then comparing equation (\ref{totder}) with the Lagrangian (\ref{eq7}) we get 
%We note here that equation (\ref{modlag}) is obtained by replacing $f_i$ by $f_{ix}$ and $g_i$ by $f_{iy}$ in (\ref{eq7}) and this is followed hereafter.  
Similarly, the condition (\ref{condition}) reduces to 
\begin{eqnarray}
\rho_{ix}=\frac{\rho_{iy}h_{ix}}{h_{iy}}. \label{frac}
\end{eqnarray}
Equation (\ref{frac}) can also be written as
\begin{eqnarray}
\left|
\begin{array}{ccc}
\rho_{ix}&\rho_{iy}\\
h_{ix}&h_{iy}
\end{array}
\right|=0.
\end{eqnarray}
Consequently the term $h_{i}$ is functionally dependent on $\rho_{i}$ that is $h_{i}=Q_{i}(\rho_{i})$.\\
Then the Lagrangian (\ref{modlag}) can also be written as 
\begin{eqnarray}
L=\sum_{i=1}^{2}\frac{1}{(\rho_{ix}\dot x+\rho_{iy}\dot y+Q_{i}(\rho_{i}))}.\label{rearr1}
\end{eqnarray}
~~~~~~~~Now the modified Emden equation is a special case of  Li\'enard type of oscillators \cite{CSL:02} which is obtained for specific forms of $Q_{i}(\rho_{i})=\rho_{i}^2+\lambda_{i}$, where $\lambda_{i}'s$ are constants. \\
From the Lagrangian (\ref{rearr1}), the corresponding equation of motion is 
%%%%%right\begin{subequations}
\begin{eqnarray}
\addtocounter{equation}{-1}
\label{equation12}
\addtocounter{equation}{1}
&&\ddot x=\bigg((\rho_{1xx}\rho_{2y}-\rho_{1y}\rho_{2xx})\dot x^2+(\rho_{1yy}\rho_{2y}-\rho_{1y}\rho_{2yy})\dot y^2+3(\rho_{1}\rho_{1x}\rho_{2y}\nonumber\\
&&\hspace{0.5cm}-\rho_{2}\rho_{1y}\rho_{2x})\dot x+3 \rho_{1y}\rho_{2y}(\rho_{1}-\rho_{2})\dot y+2(\rho_{1xy}\rho_{2y}-\rho_{1y}\rho_{2xy})\dot x\dot y\nonumber\\
&&\hspace{0.5cm}+\rho_{1}\rho_{2y}(\rho_{1}^2+\lambda_{1})-\rho_{2}\rho_{1y}(\rho_{2}^2+\lambda_{2})\bigg)(\rho_{1y}\rho_{2x}-\rho_{1x}\rho_{2y})^{-1},\label{equation1}\\
%\end{eqnarray}
%\begin{eqnarray}
&&\ddot y=\bigg( (\rho_{1x}\rho_{2xx}-\rho_{1xx}\rho_{2x})\dot x^{2}+(\rho_{1x}\rho_{2yy}-\rho_{1yy}\rho_{2x})\dot y^{2}+3\rho_{1x}\rho_{2x}(\rho_{2}\nonumber\\
&&\hspace{0.5cm}-\rho_{1})\dot x+3(\rho_{2}\rho_{1x}\rho_{2y}-\rho_{1}\rho_{1y}\rho_{2x})\dot y+2(\rho_{1x}\rho_{2xy}-\rho_{2x}\rho_{1xy})\dot x\dot y\nonumber\\
&&\hspace{0.5cm}+\rho_{2}\rho_{1x}(\rho_{2}^2+\lambda_{2})-\rho_{1}\rho_{2x}(\rho_{1}^2+\lambda_{1})\bigg)(\rho_{1y}\rho_{2x}-\rho_{1x}\rho_{2y})^{-1}.\label{equation2}
\end{eqnarray}
%%%%%%%%right\end{subequations}
The associated Hamiltonian becomes
\begin{eqnarray}
&&H=\bigg((\rho_{2y}p_{1}-\rho_{2x}p_{2})(\rho_{1}^2+\lambda_{1}) +(\rho_{1x}p_{2}-\rho_{1y}p_{1})(\rho_{2}^2+\lambda_{2})\nonumber\\
&&\hspace{0.5cm} -2\sqrt{(\rho_{1y}\rho_{2x}-\rho_{1x}\rho_{2y})}(\sqrt{(\rho_{1x}p_{2}-\rho_{1y}p_{1})}+\sqrt{(\rho_{2y}p_{1}-\rho_{2x}p_{2})}) \bigg)\nonumber\\
&&\hspace{0.5cm}\times(\rho_{1y}\rho_{2x}-\rho_{1x}\rho_{2y})^{-1}.\label{arbham}
\end{eqnarray}
Here the conjugate momenta $p_{1}$ and $p_{2}$ are defined as
%%%%%%\right\begin{subequations}
\begin{eqnarray}
&&\hspace{-1cm}p_{1}=L_{\dot x}=-\frac{\rho_{1x}}{(\rho_{1x}\dot x+\rho_{1y}\dot y+\rho_{1}^2+\lambda_{1})^2}-\frac{\rho_{2x}}{(\rho_{2x}\dot x+\rho_{2y}\dot y+\rho_{2}^2+\lambda_{2})^2},\\
&&\hspace{-1cm}p_{2}=L_{\dot y}=-\frac{\rho_{1y}}{(\rho_{1x}\dot x+\rho_{1y}\dot y+\rho_{1}^2+\lambda_{1})^2}-\frac{\rho_{2y}}{(\rho_{2x}\dot x+\rho_{2y}\dot y+\rho_{2}^2+\lambda_{2})^2}.
\end{eqnarray}
%%%%%%%%%%%right\end{subequations}
The Hamiltonian (\ref{arbham}) is connected to the Hamiltonian of  a system of uncoupled  linear harmonic oscillators 
 $\tilde{H}=\frac{1}{2}(P_{1}^2+P_{2}^2+\lambda_{1}U_{1}^2+\lambda_{2}U_{2}^2)$ through the following canonical transformation,

\begin{eqnarray}
&&P_1=\left[\lambda_1+\left(\lambda_1^2-2\lambda_1\left[\frac{p_1\rho_{2y}-p_2\rho_{2x}}{\rho_{2y}\rho_{1x}-\rho_{1y}\rho_{2x}}\right]\right)^{\frac{1}{2}}\right],\label{xx}\\
&&P_2=\left[\lambda_2+\left(\lambda_2^2-2\lambda_2\left[\frac{p_2\rho_{1x}-p_1\rho_{1y}}{\rho_{2y}\rho_{1x}-\rho_{1y}\rho_{2x}}\right]\right)^{\frac{1}{2}}\right],\\
&&U_1=-\frac{\rho_1}{\lambda_1}\left[\lambda_1^2-2\lambda_1\left(\frac{p_1\rho_{2y}-p_2\rho_{2x}}{\rho_{2y}\rho_{1x}-\rho_{1y}\rho_{2x}}\right)\right]^{\frac{1}{2}},\\
&&U_2=-\frac{\rho_2}{\lambda_2}\left[\lambda_2^2-2\lambda_2\left(\frac{p_2\rho_{1x}-p_1\rho_{1y}}{\rho_{2y}\rho_{1x}-\rho_{1y}\rho_{2x}}\right)\right]^{\frac{1}{2}}.\label{yy}
\end{eqnarray}

The general solution of (\ref{equation1}) and (\ref{equation2}) can be found after choosing the forms of $f_{1}$ and $f_{2}$ by using the relations (\ref{xx}) -(\ref{yy}) and the harmonic oscillator solution
\begin{eqnarray}
U_{i}=A_{i}\sin(\omega_{i}t+\delta_{i})\nonumber\\
P_{i}=A_{i}\omega_{i}\cos(\omega_{i}t+\delta_{i}),\quad \omega_i=\sqrt{\lambda_i}
\end{eqnarray}
where $A_{i}$ and $\delta_{i}$, ($i=1,2,$) are arbitrary constants. The above canonical transformations are identified by generalizing the knowledge of the canonical transformation for the scalar equation which is discussed in the appendix.

\subsection{ Nonlocal transformation}
The system of coupled mixed MEE equations (\ref{equation1}) and (\ref{equation2}) can also be related to a system of uncoupled simple harmonic oscillators,
\begin{eqnarray}
\ddot{u}+\lambda_1u=0,\qquad \ddot{v}+\lambda_2v=0,\label{sho}
\end{eqnarray}
through the nonlocal transformations,
\begin{eqnarray}
u=\rho_1(x,y)e^{\int \rho_1(x,y)dt},\qquad v=\rho_2(x,y)e^{\int \rho_2(x,y)dt}.\label{nonlocal}
\end{eqnarray}
Equations having such type of nonlocal transformations and their solutions have been studied in Ref. \cite{nonlocal-connection}.  In addition to using the canonical transformation obtained in the previous subsection to find the solution of equation (\ref{equation12}), one can also obtain the solution of it by solving the following set of coupled first order equations arising from (\ref{sho}) and (\ref{nonlocal}), 
%%%%%%%right\begin{subequations}
\begin{eqnarray}
\addtocounter{equation}{-1}
\label{coupled-eq1}
\addtocounter{equation}{1}
\dot{x}=\frac{uv(\rho_2^2\rho_{1y}-\rho_1^2\rho_{2y})+\dot{u}v\rho_1\rho_{2y}-u\dot{v}\rho_2\rho_{1y}}{uv(\rho_{1x}\rho_{2y}-\rho_{1y}\rho_{2x})},\\
\dot{y}=\frac{uv(\rho_1^2\rho_{2x}-\rho_2^2\rho_{1x})+u\dot{v}\rho_2\rho_{1x}-\dot{u}v\rho_1\rho_{2x}}{uv(\rho_{1x}\rho_{2y}-\rho_{1y}\rho_{2x})}\label{coupled-eq2}.
\end{eqnarray}
%%%%%%%%right\end{subequations}
Here $u$ and $v$ are the solutions of the simple harmonic oscillator equations (\ref{sho}). The difficulty in solving the equations (\ref{coupled-eq1}) depends on the form of $\rho_1$ and $\rho_2$ (see \cite{nonlocal-connection}).

%%%%%%%%%%%%%%%%%%%%%%%%%%%%%%%%%%%%%%%%%%%%%%%%%%%%%%%%%%%%%%%%%%%%%%%%%%%%%%%%%%%%%%%%%%%%%%
\section{\bf{A system of coupled Li\'enard-type equations}}
 Next we can obtain a special case of Li\'enard-type of oscillators (\ref{rearr1}) by choosing 
\begin{eqnarray}
\rho_{1}=k_{1}(\alpha_{1}x^{m_{1}}+\alpha_{2}y^{m_{2}}), \quad \rho_{2}=k_{2}(\alpha_{3}y^{m_{3}}+\alpha_{4}x^{m_{4}}). 
\end{eqnarray}
where $k_{i}$'s and $\alpha_{j}$'s are arbitrary real parameters and $m_{j}$'s are positive integers and $i=1,2,$ and $j=1,2,3,4.$
  In this case the specific form of the nonstandard and nonsingular Lagrangian with two degrees of freedom takes the form
\begin{eqnarray}
&&\hspace{-1.3cm}L=\frac{1}{\lambda_{1}+k_{1}^2(\alpha_{1}x^{m_{1}}+\alpha_{2}y^{m_{2}})^2+k_{1}(m_{1}\alpha_{1}x^{m_{1}-1}\dot x+m_{2}\alpha_{2}y^{m_{2}-1}\dot y)}\nonumber\\
&&\hspace{-.5cm}+\frac{1}{\lambda_{2}+k_{2}^2(\alpha_{3}y^{m_{3}}+\alpha_{4}x^{m_{4}})^2+k_{2}(m_{3}\alpha_{3}y^{m_{3}-1}\dot y+m_{4}\alpha_{4}x^{m_{4}-1}\dot x)}.\label{lagpowerm}
\end{eqnarray}
The equation of motion of the mixed Li\'enard-type class of oscillators can be obtained from the above Lagrangian.  It is of the form
%%%%%%%right\begin{subequations}
\label{secpowerm1&m2}
\begin{eqnarray}
&&\ddot x=\frac{-1}{xy^2\delta_{m}}\bigg(\dot x^2 y^2\bigg[m_{1}m_{3}\alpha_{1}\alpha_{3}(m_{1}-1)x^{m_{1}}y^{m_{3}}-m_{2}m_{4}\alpha_{2}\alpha_{4}(m_{4}-1)x^{m_{4}}y^{m_{2}}  \bigg]\nonumber\\
&&+
\dot y^2 x^2\bigg[    m_{2} m_{3} \alpha_{2}\alpha_{3}(m_{2}-m_{3})y^{m_{2}+m_{3}}\bigg]+ \bigg[3k_{1}y^{m_{3}} (m_{1}\alpha_{1}x^{m_{1}}y\dot x+m_{2}\alpha_{2}y^{m_{2}}x\dot y)\nonumber\\
&& +k_{1}^2xy^{m_{3}+1}(\alpha_{1} x^{m_{1}}+\alpha_{2} y^{m_{2}})^2+\lambda_{1}x y^{m_{3}+1} \bigg]m_{3} \alpha_{3} x y (\alpha_{1} x^{m_{1}}+\alpha_{2} y^{m_{2}})\nonumber\\
&&-\bigg[3k_{2}y^{m_{2}} (m_{4}\alpha_{4}x^{m_{4}}y\dot x+m_{3}\alpha_{3}y^{m_{3}}x\dot y) +k_{2}^2xy^{m_{2}+1}(\alpha_{4} x^{m_{4}}+\alpha_{3} y^{m_{3}})^2\nonumber\\
&&+\lambda_{2}x y^{m_{2}+1} \bigg]m_{2} \alpha_{2} x y (\alpha_{4} x^{m_{4}}+\alpha_{3} y^{m_{3}}) \bigg),\label{secpowerm1}
\end{eqnarray}

\begin{eqnarray}
&&\ddot y=\frac{1}{yx^2\delta_{m}}\bigg(\dot y^2x^2\bigg[m_{2}m_{4}\alpha_{2}\alpha_{4}(m_{2}-1)x^{m_{4}}y^{m_{2}} -m_{1}m_{3}\alpha_{1}\alpha_{3}(m_{3}-1)x^{m_{1}}y^{m_{3}}\bigg]\nonumber\\
&&-\dot x^2y^{2}\bigg[m_{1}m_{4}\alpha_{1}\alpha_{4}(m_{4}-m_{1})x^{m_{1}+m_{4}} \bigg] 
+ \bigg[3k_{1} x^{m_{4}}(m_{1}\alpha_{1}x^{m_{1}}y\dot x+m_{2}\alpha_{2}y^{m_{2}}x\dot y)\nonumber\\
&& +k_{1}^2x^{m_{4}+1}y(\alpha_{1} x^{m_{1}}+\alpha_{2} y^{m_{2}})^2+\lambda_{1}yx^{m_{4}+1}  \bigg]m_{4} \alpha_{4} x y (\alpha_{1} x^{m_{1}}+\alpha_{2} y^{m_{2}})
\nonumber\\
&&-\bigg[3k_{2}x^{m_{1}} (m_{4}\alpha_{4}x^{m_{4}}y\dot x+m_{3}\alpha_{3}y^{m_{3}}x\dot y) +k_{2}^2x^{m_{1}+1}y(\alpha_{4} x^{m_{4}}+\alpha_{3} y^{m_{3}})^2\nonumber\\
&&+\lambda_{2}x^{m_{1}+1} y \bigg]m_{1} \alpha_{1} x y (\alpha_{4} x^{m_{4}}+\alpha_{3} y^{m_{3}}) \bigg),\label{secpowerm2}
\end{eqnarray}
%%%%%%right\end{subequations}

where $\delta_{m}=m_{1}m_{3}\alpha_{1}\alpha_{3}x^{m_{1}}y^{m_{3}}-m_{2}m_{4}\alpha_{2}\alpha_{4}x^{m_{4}}y^{m_{2}}$. \\

\noindent The associated Hamiltonian becomes
\begin{eqnarray}
&&\hspace{-0.3cm}H=\frac{1}{k_{1}k_{2}\delta_{m}}\bigg[k_{2}(m_{4}\alpha_{4}p_{2}x^{m_{4}}y-m_{3}\alpha_{3}p_{1}xy^{m_{3}})\bigg(k_{1}^2 X_{m}^2+\lambda_{1}  \bigg)\nonumber\\
&&\hspace{0.3cm}+k_{1}(m_{2}\alpha_{2}p_{1}xy^{m_{2}}-m_{1}\alpha_{1}p_{2}x^{m_{1}}y)\bigg(k_{2}^2 Y_{m}^2+\lambda_{2}  \bigg)-2\delta_{m}^{\frac{1}{2}}\bigg(k_{1}k_{2}^{\frac{1}{2}}(m_{2}\alpha_{2}p_{1}xy^{m_{2}}\nonumber\\
&&\hspace{0.5cm}-m_{1}\alpha_{1}p_{2}x^{m_{1}}y)^{\frac{1}{2}}+k_{2}k_{1}^{\frac{1}{2}}(m_{4}\alpha_{4}p_{2}x^{m_{4}}y-m_{3}\alpha_{3}p_{1}xy^{m_{3}})^{\frac{1}{2}}  \bigg)\bigg].
\end{eqnarray}
where $X_{m}=\alpha_{1}x^{m_{1}}+\alpha_{2}y^{m_{2}}$, $Y_{m}=\alpha_{4}x^{m_{4}}+\alpha_{3}y^{m_{3}}$.\\
%%%%%%%%%%%%%%%%%%%%%%%%%%%%%%%%%%%%%%%%%%%%%%%%%%%%%%%%%%%%%%%%%%%%%%%%%%%%%%%%%%%%%%%%%%%%
\subsection{\bf{Coupled Modified Emden Equation: A special case}}
	The coupled system of equations (\ref{secpowerm1}) and (\ref{secpowerm2}) reduces to a coupled generalization of the modified Emden equation (\ref{meee})-(\ref{mee}) for the choice $m_{1}$ = $m_{2}$ = $m_{3}$ = $m_{4}$ = $1$ and is of the form 
%%%right\begin{subequations} 
\begin{eqnarray}
\addtocounter{equation}{-1}
%\label{eq15and15a}
\addtocounter{equation}{1}
&\hspace{-1cm}\ddot x=\frac{-1}{\hat{\delta_{1}}}\bigg[\bigg(3 k_{1}(\alpha_{1}\dot x+\alpha_{2}\dot y)+k_{1}^2(\alpha_{1}x+\alpha_{2}y)^2 +\lambda_{1}\bigg)(\alpha_{3}(\alpha_{1}x+\alpha_{2}y))\nonumber\\
&\hspace{-1cm}-\bigg(3 k_{2}(\alpha_{3}\dot y+\alpha_{4}\dot x)+k_{2}^2(\alpha_{3}y+\alpha_{4}x)^2 +\lambda_{2}\bigg)(\alpha_{2}(\alpha_{3}y+\alpha_{4}x)) \bigg],\label{eq15}
\end{eqnarray}
\begin{eqnarray}
&\hspace{-1cm}\ddot y=\frac{1}{\hat{\delta_{1}}}\bigg[\bigg(3 k_{1}(\alpha_{1}\dot x+\alpha_{2}\dot y)+k_{1}^2(\alpha_{1}x+\alpha_{2}y)^2 +\lambda_{1}\bigg)(\alpha_{4}(\alpha_{1}x+\alpha_{2}y))\nonumber\\&\hspace{-1cm}-\bigg(3 k_{2}(\alpha_{3}\dot y+\alpha_{4}\dot x)+k_{2}^2(\alpha_{3}y+\alpha_{4}x)^2 +\lambda_{2}\bigg)(\alpha_{1}(\alpha_{3}y+\alpha_{4}x)) \bigg],\label{eq15a}
\end{eqnarray}
%%%%%%%%right\end{subequations} 
where $\hat{\delta_{1}}=(\alpha_{1}\alpha_{3}-\alpha_{2}\alpha_{4})$.  Note that the above system of equations is \emph{PT} symmetric under the combined transformations, $(t\rightarrow-t$, $x\rightarrow-x$, $ y\rightarrow- y)$.  There are other ways of generalizing the modified Emden equation to two dimensions, for example see Ref. \cite{cmee}.  Even though the generalized version identified as the coupled modified Emden equation in Ref. \cite{cmee} is isochronous, the system lacks a Hamiltonian description in order to quantize it.  However, we find that the system (\ref{eq15}) and (\ref{eq15a}) has the well defined Hamiltonian 
\begin{eqnarray}
&&\hspace{-0.7cm}H=\frac{1}{k_{1}k_{2}\hat{\delta_{1}}}\bigg[k_{2}(p_{2}\alpha_{4}-p_{1}\alpha_{3})(k_{1}^2 X^2+\lambda_{1})+k_{1}(p_{1}\alpha_{2}-p_{2}\alpha_{1})(k_{2}^2 Y^2+\lambda_{2})\nonumber\\
&&\hspace{-0.1cm}  -2\sqrt{\hat{\delta_{1}}}(k_{1}k_{2}^{\frac{1}{2}}(p_{1}\alpha_{2}-p_{2}\alpha_{1})^{\frac{1}{2}}+   k_{2}k_{1}^{\frac{1}{2}}(p_{2}\alpha_{4}-p_{1}\alpha_{3})^{\frac{1}{2}}) \bigg],\label{eq16}
\end{eqnarray}
where $X=(\alpha_{1}x+\alpha_{2}y)$ and $Y=(\alpha_{3}y+\alpha_{4}x)$.   Here we note that one can use a variable transformation $X=(\alpha_{1}x+\alpha_{2}y)$ and $Y=(\alpha_{3}y+\alpha_{4}x)$ in equation (\ref{eq15}) to obtain two uncoupled modified Emden equations. 

In order to find the general solution of the equations (\ref{eq15}) and (\ref{eq15a}) we use suitable canonical transformation to the above Hamiltonian to reduce it to a simpler form.  We find that the above Hamiltonian is connected to the Hamiltonian of a two dimensional linear harmonic oscillator through the following canonical transformation, see (\ref{xx})-(\ref{yy}),	

\begin{eqnarray}
&&P_1=\lambda_1+\left[\lambda_1^2-\frac{2\lambda_1(\alpha_3p_1-\alpha_4p_2)}{k_1(\alpha_1\alpha_3-\alpha_2\alpha_4)}\right]^{\frac{1}{2}},\label{eq119}\\
&&P_2=\lambda_2+\left[\lambda_2^2-\frac{2\lambda_2(\alpha_1p_2-\alpha_2p_1)}{k_2(\alpha_1\alpha_3-\alpha_2\alpha_4)}\right]^{\frac{1}{2}},\\
&&U_1=-\frac{k_1(\alpha_1x+\alpha_2y)}{\lambda_1}\left[\lambda_1^2-\frac{2\lambda_1(\alpha_3p_1-\alpha_4p_2)}{k_1(\alpha_1\alpha_3-\alpha_2\alpha_4)}\right]^{\frac{1}{2}},\\
&&U_2=-\frac{k_2(\alpha_4x+\alpha_3y)}{\lambda_2}\left[\lambda_2^2-\frac{2\lambda_2(\alpha_1p_2-\alpha_2p_1)}{k_2(\alpha_1\alpha_3-\alpha_2\alpha_4)}\right]^{\frac{1}{2}}\label{eq120},
\end{eqnarray}
\begin{figure}
\centering 
\includegraphics[width=1.1\columnwidth]{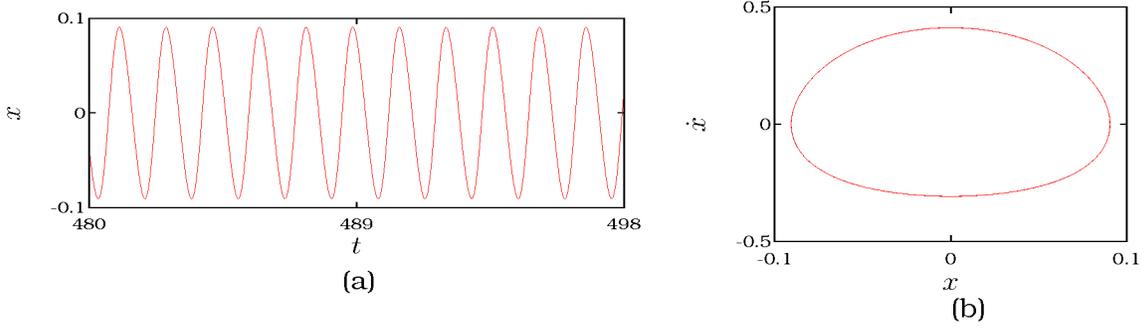}
\caption{(color online) Periodic oscillations with $\omega_{1}:\omega_{2}$ = $4:4$, $k_{1}$ = $k_{2}$ = 1,  $\alpha_{1}$ = $\alpha_{3}$ = $5.5$, $\alpha_{2}$ = $\alpha_{4}$ = $3$ (a) Time series plot (b) Projected phase portrait.  Similar plots can be given for the $y$ variable.}
\end{figure}
\begin{figure}
\centering 
\includegraphics[width=1.0\columnwidth]{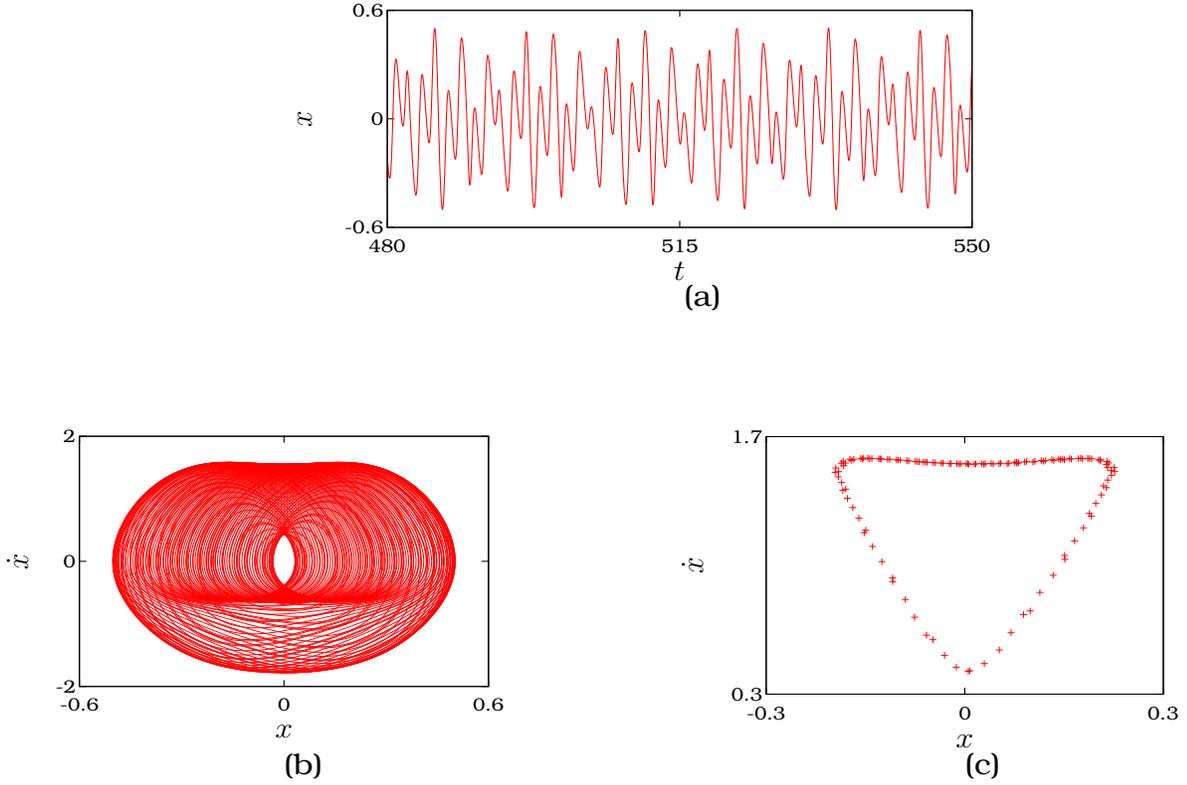}
\caption{(color online) Quasiperiodic oscillations with $\omega_{1}:\omega_{2}$ = $4$ : $\sqrt{3}$, $k_{1}$ = $k_{2}$ = $1$,  $\alpha_{1}$ = $\alpha_{3}$ = $5.5$, $\alpha_{2}$ = $\alpha_{4}$ = $3$. (a) Time series plot (b) Projected phase portrait (c) Poincar\'e section.  Similar plots can be given for the $y$ variable.}
\end{figure}

The general solution of equations (\ref{eq15}) and (\ref{eq15a}) can then be obtained by substituting the general solution of the two dimensional harmonic oscillator given by the expressions
\begin{eqnarray}
U_{1}=A \sin(\omega_{1} t+ \delta_{1}),\quad U_{2}=B \sin(\omega_{2} t+ \delta_{2}),\nonumber \\
P_{1}=A \omega_{1}\cos(\omega_{1}t+\delta_{1}),\quad P_{2}=B \omega_{2}\cos(\omega_{2}t+\delta_{2}),\label{harmq}
\end{eqnarray}
where $\omega_{j}=\sqrt{\lambda_j}$, j=1,2 into Eqs. (\ref{eq119})-(\ref{eq120}) and solving the resultant equations.  We obtain
%%%%%%%right\begin{subequations}
\begin{eqnarray}
x=\frac{A\alpha_{3}\omega_{1}\sin(\omega_{1}t+\delta_{1})}{k_{1}\hat{\delta_{1}}\left(\omega_{1}-A\cos(\omega_{1}t+\delta_{1})\right)}-\frac{B\alpha_{2}\omega_{2}\sin(\omega_{2}t+\delta_{2})}{k_{2}\hat{\delta_{1}}\left(\omega_{2}-B\cos(\omega_{2}t+\delta_{2})\right)},
\end{eqnarray}
\begin{eqnarray}
y=-\frac{A\alpha_{4}\omega_{1}\sin(\omega_{1}t+\delta_{1})}{k_{1}\hat{\delta_{1}}\left(\omega_{1}-A\cos(\omega_{1}t+\delta_{1})\right)}+\frac{B\alpha_{1}\omega_{2}\sin(\omega_{2}t+\delta_{2})}{k_{2}\hat{\delta_{1}}\left(\omega_{2}-B\cos(\omega_{2}t+\delta_{2})\right)},
\end{eqnarray}\\
%%%%%right\end{subequations}
where $A$,\,$B$,\,$\delta_{1}$, $\delta_{2}$ are arbitrary constants. 
The above solution is oscillatory and is periodic and bounded for suitable choice of parameters namely $0<A<\omega_{1}$ and $0<B<\omega_{2}$.
Here one can note that the frequency of oscillations is again independent of the amplitude for the two dimensional coupled modified Emden equation.  Figures 1 and 2 show two types of oscillatory behaviour, namely periodic and quasiperiodic oscillations, depending upon the ratio of frequencies $\omega_{1}$ and $\omega_{2}$.  Fig.1 shows  periodic oscillations for $\omega_1$ : $\omega_2$ = $4:4$, $k_{1}$ = $k_{2}$ = $1$, $\alpha_{1}$ = $\alpha_{3}$ = $5.5$, $\alpha_{2}$ = $\alpha_{4}$ = $3$ and Fig.2 shows quasiperiodic oscillations for $\omega_1$ : $\omega_2$ = $4$ : $\sqrt{3}$, $k_{1}$ = $k_{2}$ = $1$, $\alpha_{1}$ = $\alpha_{3}$ = $5.5$, $\alpha_{2}$ = $\alpha_{4}$ = $3$.

%\begin{eqnarray}
%&&\hspace{-2cm}\ddot x=\frac{1}{(\alpha_{1}\alpha_{3}-\alpha_{2}\alpha_{4})}\bigg[3(k_{2}\alpha_{2}\alpha_{4}^2-k_{1}\alpha_{3}\alpha_{1}^2)x\dot x+3\alpha_{2}\alpha_{3}\bigg((k_{2}\alpha_{3}-k_{1}\alpha_{2})y\dot y\nonumber\\&&\hspace{-1cm}+(k_{2}\alpha_{4}-k_{1}\alpha_{1})(y\dot x +x  \dot y)\bigg)   +  (k_{2}^2\alpha_{4}^3\alpha_{2}-k_{1}^2\alpha_{1}^3\alpha_{3})x^3+\alpha_{2}\alpha_{3}\bigg((k_{2}^2\alpha_{3}^2-k_{1}^2\alpha_{2}^2)y^3 \nonumber\\&&\hspace{-1cm}+3(k_{2}^2\alpha_{4}^2-k_{1}^2\alpha_{1}^2)x^{2}y+ 3(k_{2}^2\alpha_{3}\alpha_{4}-k_{1}^2\alpha_{1}\alpha_{2})xy^{2}+(\lambda_{2}-\lambda_{1})y\bigg)\nonumber\\&&\hspace{-1cm}+(\alpha_{2}\alpha_{4}\lambda_{2}-\alpha_{1}\alpha_{3}\lambda_{1})x\bigg],\label{eq15}
%\end{eqnarray}
%\begin{eqnarray}
%&&\hspace{-2cm}\ddot y=\frac{1}{(\alpha_{1}\alpha_{3}-\alpha_{2}\alpha_{4})}\bigg[3\alpha_{1}\alpha_{4}\bigg((k_{1}\alpha_{1}-k_{2}\alpha_{4})x\dot x+(k_{1}\alpha_{2}-k_{2}\alpha_{3})(y\dot x +x  \dot y)\bigg)\nonumber\\&&\hspace{-1cm} +3(k_{1}\alpha_{4}\alpha_{2}^2-k_{2}\alpha_{1}\alpha_{3}^2)y\dot y+\alpha_{1}\alpha_{4}\bigg((k_{1}^2\alpha_{1}^2-k_{2}^2\alpha_{4}^2)x^3+3(k_{1}^2\alpha_{1}\alpha_{2}-k_{2}^2\alpha_{3}\alpha_{4})x^{2}y\nonumber\\&&\hspace{-1cm}+3(k_{1}^2\alpha_{2}^2-k_{2}^2\alpha_{3}^2)xy^{2}+ (\lambda_{1}-\lambda_{2})x\bigg)+(k_{1}^2\alpha_{2}^3\alpha_{4}-k_{2}^2\alpha_{3}^3\alpha_{1})y^3 \nonumber\\&&\hspace{-1cm}+(\alpha_{2}\alpha_{4}\lambda_{1}-\alpha_{1}\alpha_{3}\lambda_{2})y\bigg],\label{eq15a}
%\end{eqnarray}

\section{Conclusion}
In this paper we have identified a class of coupled mixed Li\'enard type nonlinear evolution equations.  This class of equation is identified by generalizing the nonstandard Lagrangian of the scalar MEE to a suitable nonsingular two dimensional Lagrangian.  Imposing the condition that the resultant Euler-Lagrange equation should be of mixed Li\'enard type form, we have identified a specific class of equations which admits isochoronous solutions. The procedure is illustrated for the special case of  two dimensional modified Emden equation and is found to be isochronous as well as $\emph PT$ symmetric for suitable choice of parameters, exhibiting quasiperiodic and periodic oscillations.  The above procedure can in principle be extended to higher degrees of freedom also.  The problem of quantization of system  (\ref{eq16}) is also worth further consideration.

\section{Acknowledgements}
The work of ADD and ML is supported by a DST-IRHPA research project and the work of ML is also supported by a DAE Raja Ramanna Fellowship.   The work of V.K.C. is supported by the SERB-DST Fast Track scheme for young scientists under Grant No. YSS/2014/000175.

\section*{Appendix}

Let us consider the following mixed Li\'enard type equation,
\begin{eqnarray}
\ddot{x}+\dot{x}^2\frac{u_{xx}}{u_x}+3ku\dot{x}+
k^2\frac{u^3}{u_x}+\lambda\frac{u}{u_x}=0,
\end{eqnarray}
where $u(x)$ is an arbitrary function of $x$.  This equation is obtained by applying the transformation $y=u(x)$ in the linearizable MEE
\begin{eqnarray}
\ddot{y}+3ky\dot{y}+k^2y^3+\lambda y=0.
\end{eqnarray}
The resultant equation
possess the Hamiltonian structure
\begin{eqnarray}
H=-kp\frac{u^2}{u_x}-\frac{\lambda p}{k u_{x}}+2\sqrt{-\frac{p}{k u_x}},
\end{eqnarray}
which can be deduced from the Lagrangian (\ref{arblag}).  This Hamiltonian can be transformed to the Hamiltonian of the simple
harmonic oscillator equation through the canonical transformation
\begin{eqnarray}
U=\frac{u(x)\sqrt{\lambda(\lambda-\frac{2kp}{u_x})}}{\lambda},\qquad P=\frac{\lambda-\sqrt{\lambda(\lambda-\frac{2kp}{u_x})}}{k}.
\end{eqnarray}
Generalizing the above canonical transformation to two dimensions we obtain the canonical transformations (vide Eqs. (\ref{xx})-(\ref{yy})) for the coupled MEE.\\
\section*{References}

%\begin{acknowledgements}
%\end{acknowledgements}

%\bibliography{reference}   % name your BibTeX data base

%\begin{thebibliography}{}
% etc
%\end{thebibliography}

\end{document}